\documentclass[letter,traditabstract]{aa}
\usepackage{graphicx}
\usepackage{txfonts}
\usepackage{natbib} 
\bibpunct{(}{)}{;}{a}{}{,} 

\begin{document}

\title{3D simulations of supernova remnants evolution\\ including non-linear particle acceleration}
\titlerunning{3D simulations of supernova remnants evolution including non-linear particle acceleration}

\author{Gilles Ferrand\inst{1}, Anne Decourchelle\inst{1}, Jean Ballet\inst{1}\\
Romain Teyssier\inst{1,2}, Federico Fraschetti\inst{3,4}}
\authorrunning{Ferrand et al}
\institute{
Laboratoire AIM (CEA/Irfu, CNRS/INSU, Universit{\'e} Paris VII), CEA Saclay, b{\^a}t. 709, F-91191 Gif sur Yvette, France
\and
Institute for Theoretical Physics, University of Z{\"u}rich, CH-8057 Z{\"u}rich, Switzerland
\and
LUTh, Observatoire de Paris, CNRS-UMR8102 et Universit{\'e} Paris VII, F-92195 Meudon C{\'e}dex, France
\and
Lunar and Planetary Lab \& Dept. of Physics, University of Arizona, Tucson, AZ, 85721, USA
}

\date{draft 2.8 - Dec. 22, 2009}

\abstract{
If a sizeable fraction of the energy of supernova remnant shocks is channeled 
into energetic particles (commonly identified with Galactic cosmic rays), 
then the morphological evolution of the remnants must be distinctly modified. 
Evidence of such modifications has been recently obtained with the
\emph{Chandra} and \emph{XMM-Newton} X-ray satellites. 
To investigate these effects,
we coupled a semi-analytical kinetic model of shock acceleration 
with a 3D hydrodynamic code (by means of an effective adiabatic index). 
This enables us to study the time-dependent compression 
of the region between the forward and reverse shocks due to the back reaction of accelerated particles, 
concomitantly with the development of the Rayleigh-Taylor hydrodynamic instability at the contact discontinuity.
Density profiles depend critically on the injection level~$\eta$ of particles: 
for $\eta \lesssim 10^{-4}$ modifications are weak and progressive,
for $\eta \sim 10^{-3}$ modifications are strong and immediate.
Nevertheless, the extension of the Rayleigh-Taylor unstable region does not depend on the injection rate. 
A first comparison of our simulations with observations of \emph{Tycho}'s remnant 
strengthens the case for efficient acceleration of protons at the forward shock.}

\maketitle


\section{Introduction
\label{Introduction}}

Supernova remnants (SNRs) are believed to be the major contributors
to Galactic cosmic rays. But although acceleration of electrons in these objects
is well established, direct evidence of the acceleration of protons remains elusive: 
recent detections of non-thermal gamma rays do not yet allow 
unambiguously distinguishing leptonic and hadronic contributions
(see e.g. \citealt{Gabici2008a} and references therein).

To assess particle acceleration in SNRs, a promising alternative approach 
consists in diagnosing the impact of energetic particles on the SNR dynamics.
Indeed, if SNRs are efficient accelerators, as claimed, then
energetic nuclei must make a sizable impact on their evolution 
(see e.g. \citealt{Jones1991a}, \citealt{Malkov2001c}).
Such a study is now possible thanks to the performance 
of modern X-ray observatories such as \emph{Chandra} and \emph{XMM-Newton}, 
which allow the structure of young Galactic SNRs to be probed in great detail.
\cite{Warren2005a} have been able to determine the positions of
the three waves (forward shock, contact discontinuity, reverse shock)
in \emph{Tycho}'s SNR with rather good accuracy, and have found
that it does not match any pure hydrodynamic model. The same effect
has been reported in SN~1006 (\citealt{Cassam-Chenai2008a}, \citealt{Miceli2009a})
and in Cas~A (\citealt{Patnaude2009b}).
This is evidence that not all the kinetic energy from the explosion
is converted into heat downstream of the shock, 
but that a sizeable part is channeled elsewhere
-- probably into energetic particles.

This effect has been studied by \cite{Decourchelle2000a}, using
1D self-similar simulations coupled with a simple model of non-linear acceleration
(from \citealt{Berezhko1999a}). 
They have shown how the shocked region shrinks
in the case of efficient acceleration of particles at the shocks:
as the injection fraction rises the shocks get closer to the contact discontinuity. 
These results have been confirmed and extended by 1D hydrodynamic simulations 
of radially symmetric SNRs (see \citealt{Ellison2007a} and references therein).

But the contact discontinuity between the shocked ISM and the shocked ejecta is known
to be hydrodynamically unstable: this interface is subject to the Rayleigh-Taylor instability,
as the ejecta are being decelerated by the ambient medium of lower density.
Thus the morphological study of a SNR requires a~3D (or at least~2D) modelling
(see \citealt{Chevalier1992a}, \citealt{Dwarkadas2000a}, \citealt{Wang2001a} and references therein). 
As the instability develops the contact discontinuity is profoundly modified: 
the ejecta protrude inside the shocked ISM, forming characteristic finger-like structures.
These features are particularly apparent in \emph{Tycho}'s SNR, where the ejecta
exhibit a fleecy aspect in both X-rays \citep{Warren2005a} and in radio \citep{Velazquez1998a}.
Thus, to diagnose the back reaction of energetic particles, 
one needs to take hydrodynamic instabilities into account.
To assess their impact, \cite{Blondin2001a} have made 2D and 3D hydrodynamic simulations
of a slice of SNR, mimicking the presence of energetic particles by lowering the adiabatic index 
of the fluid (uniformly in space and time).

In this work, for the first time we combine all these previous approaches;
that is, we make full 3D simulations of an SNR evolution including 
a space- and time-dependent model of acceleration and back reaction of particles,
to be able to interpret the X-ray observations.


\section{Method
\label{sec:Method}}

\subsection{Hydrodynamic evolution\label{sub:Ramses}}

As we are interested in the time-dependent and non-linear interplay 
between the SNR and the energetic particles it accelerates, 
we resort to numerical simulations.
To compute the remnant evolution, we used the code \emph{RAMSES} \citep{Teyssier2002a}.
This 3D Cartesian Eulerian code includes a second-order hydrodynamic solver, 
and implements a tree-based structure allowing for versatile adaptive mesh refinement (AMR).

\begin{figure}[th]
\begin{centering}
\includegraphics[width=8cm]{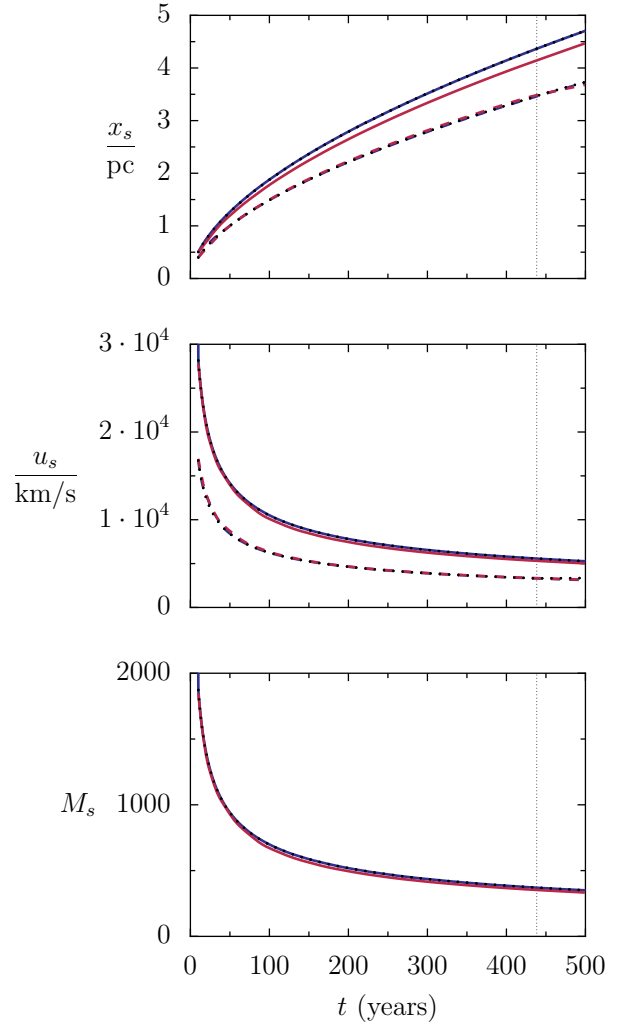}
\par\end{centering}
\caption{Evolution of the shocks.\label{fig:shocks_onoff}}
Radial position~$x_S$, velocity~$u_S$, and Mach number~$M_S$ of the forward (solid) and reverse
(dashed) shocks are plotted versus time, averaged over all directions
(Mach number is not shown for the reverse shock). 
Blue curves show the unmodified, i.e. purely hydrodynamic case
(with SNR parameters given in Sect.~\ref{sub:Chevalier}); 
red curves show the modified case, including the back reaction of accelerated particles
(with acceleration parameters given in Sect.~\ref{sub:Morphology}).
\end{figure}

Although \emph{RAMSES} has been already extensively tested and used, 
we had to adapt it to the study of supernova remnants \citep{Fraschetti2009a}. 
The main point is that we use a grid that is comoving with the contact discontinuity; 
that is, we work in an expanding frame. 
Because this frame is non-inertial, we have to modify the Euler equations. 
Although it is computationally very interesting to factor out the global expansion of the remnant in this way,
we have to face the numerical instabilities associated to quasi-stationary shock waves.
Still, we can accurately follow the SNR evolution as shown in Fig.~\ref{fig:shocks_onoff}. 
In the unmodified case, the position and speed of the shocks 
exactly coincide with analytical predictions by \cite{Truelove1999a}. 
In the simulations presented here, we actually simulate one eighth of a sphere, assuming symmetry.

\subsection{Particle acceleration and back reaction\label{sub:Blasi}}

To compute acceleration of particles by shocks, we used the 
semi-analytical model of \cite{Blasi2002a, Blasi2004a}, 
a non-linear model of diffusive shock acceleration (DSA)
that takes the back reaction of accelerated particles on the fluid structure into account.
This model solves the particle spectrum~$f(p)$ and the fluid velocity profile~$U(p)$ 
jointly as functions of the momentum~$p$ of particles.
It includes the escape of particles with the highest energy
upstream of the shock, which carry energy away from the accelerator.
We also include the effect of Alfv{\'e}n wave heating in the precursor, 
which limits shock modifications, but we do not include magnetic field amplification \citep{Amato2006a}.

As inputs, the model requires 
(i)~information on the shock
(its speed~$u_S$ and Mach number~$M_S$ are 
provided by the hydrodynamic code, averaged over the remnant surface); 
(ii)~an injection recipe (we assume that a fraction~$\eta$ of the particles crossing the shock 
enter the accelerator, at injection momentum $p_{\mathrm{inj}}$ 
equal to $\xi$~times the mean downstream thermal momentum); and 
(iii)~a cutoff recipe (we limit the maximum momentum $p_{\mathrm{max}}$
according to the age and the size of the remnant,
assuming Bohm diffusion, i.e. a diffusion coefficient 
$D(p)=D_0\:p^2/\sqrt{1+p^2}$ with $D_0=3.10^{21}\:\mathrm{cm^2/s}$).
We consider particle acceleration only at the forward shock, 
as there is less theoretical and observational evidence of efficient acceleration 
at the reverse shock (see \citealt{Ellison2005a} for a discussion of this issue).

Amongst the outputs, the acceleration model provides the total shock compression ratio~$r_{\mathrm{tot}}$.
To couple the hydrodynamic evolution of the remnant with particle acceleration, 
we vary the adiabatic index of the fluid as done in~\cite{Ellison2004a}:
at each time-step, we compute the index~$\gamma_{\mathrm{eff}}$, 
which would have produced the same ratio~$r_{\mathrm{tot}}$,
and affect it in \emph{RAMSES} to the cells located just upstream of the shock front.
Then $\gamma_{\mathrm{eff}}$ is advected inside the shocked region, 
so that it remains constant in each fluid element, 
which thus remembers modifications induced by particle acceleration at the time it was shocked.
\cite{Ellison2004a} have shown that there is good agreement between 
such a pseudo-fluid approach and two-fluid calculations in~1D.

\subsection{Supernova remnant initialisation\label{sub:Chevalier}}

We initialise our simulations at a young age (here 10~years), 
using self-similar profiles from \cite{Chevalier1983a}, 
including the pressure of accelerated particles (as computed from
the acceleration model presented in the previous section).
Assuming that both the ejecta (but for a central uniform core)
and the ambient medium have power-law density profiles (of indices respectively~$n$ and~$s$), 
hydrodynamic profiles are obtained by integration of ordinary differential equations. 

Here we are interested in a \emph{Tycho}-like SNR, that is a supernova of
type~Ia, referring to $10^{51}$~erg of kinetic energy
in 1.4~solar masses, with an $n=7$ power-law distribution. 
We assume a uniform ($s=0$) and tenuous ($n_{H,0}=0.1\,\mathrm{cm^{-3}}$) ambient medium.

Finally, we mention that Rayleigh-Taylor instabilities are not explicitly seeded in the simulation,
but are spontaneously triggered at the contact discontinuity by numerical perturbations seeded by the grid.


\begin{figure}[th]
\begin{centering}
\includegraphics[width=7.8cm]{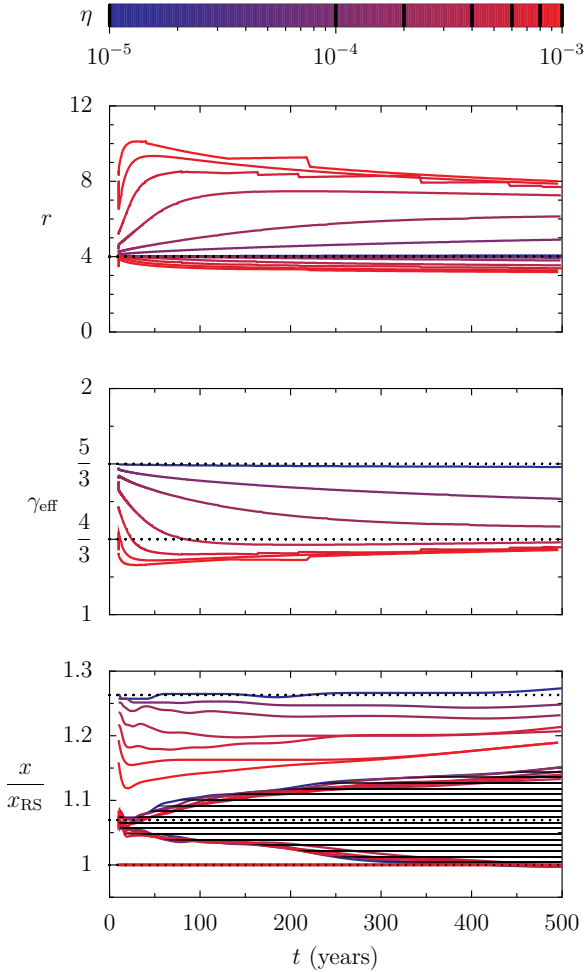}
\par\end{centering}
\caption{Evolution of the remnant structure.\label{fig:accelRS_eta}}
Top plot: compression factor~$r$ of the shock 
(upper curves show the total compression~$r_{\mathrm{tot}}$,
lower curves show the subshock compression~$r_{\mathrm{sub}}$). 
Middle plot: effective adiabatic index~$\gamma_{\mathrm{eff}}$ at the shock front. 
Bottom plot: relative positions~$x$ of the waves (averaged over all directions). 
From top to bottom, curves correspond to the forward shock, the mixing zone boundaries
(defined as the region where at least 10\% of the density is made by
a constituent which would not have come here without instabilities),
and the reverse shock. 
Colour codes the (constant) injection fraction~$\eta$, 
rising from blue (almost unmodified case) to red (very modified case) 
as follows: $10^{-5},10^{-4},2\times 10^{-4},4\times 10^{-4},6\times 10^{-4},8\times 10^{-4},10^{-3}$.
\end{figure}

\begin{figure}[th]
\begin{centering}
\includegraphics[width=8.5cm]{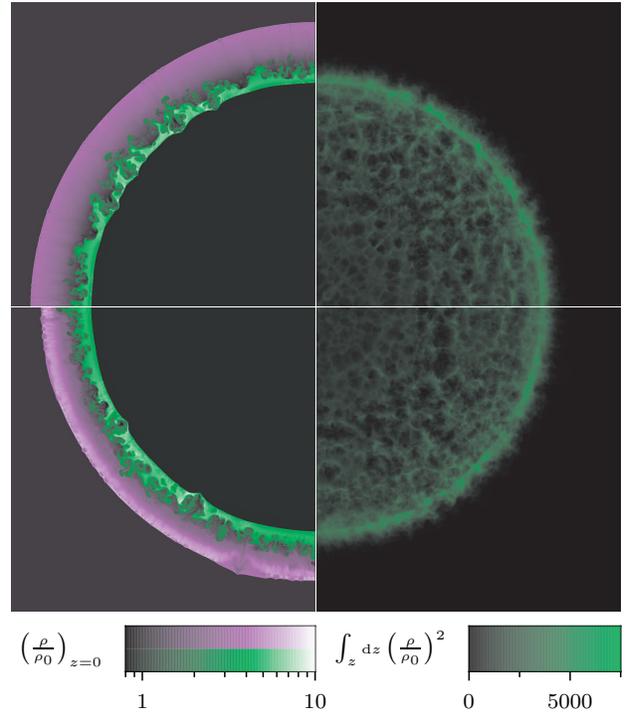}
\end{centering}
\caption{Maps of the density.\label{fig:map_onoff}}
Snapshot at time 500~years from 3D simulations with formal resolution of $1024^{3}$~cells. 
Left side: slices of the density in the plane $z=0$.
Colour codes phases: ejecta are in green (where their fraction is at least 10\%),
ambient medium is in purple ($\rho_0$ is its unperturbed mass density).
Features located just behind the forward shock result from a numerical instability.
Right side: projection of the square of the ejecta density along $z$-axis. 
Top half of the figure shows a purely hydrodynamic case,
bottom half a case including the back reaction of accelerated particles
(with the injection recipe of \citealt{Blasi2005a} which gives $\eta \sim 4.10^{-4}$).
\end{figure}

\section{Results
\label{sec:Results}}

\subsection{Remnant evolution\label{sub:Evolution}}

The temporal evolution of some key parameters is shown in Fig.~\ref{fig:accelRS_eta}.
The acceleration model depicted in Sect.~\ref{sub:Blasi} 
provides the compression ratios plotted on top of the figure.
For ease of interpretation, we ran multiple simulations with 
different fixed injection rates~$\eta$. If $\eta \simeq 10^{-5}$ (in dark blue), 
the system is almost in the linear (test-particle) regime, and 
there is a single strong shock of compression ratio $r=4$. 
As we raise~$\eta$ to $10^{-3}$ (as colour gets warmer), 
it progressively enters the non-linear (modified) regime, 
and the shock discontinuity is reduced to a subshock of compression~$r_{\mathrm{sub}}$
between~3 and~4, whereas the overall compression ratio~$r_{\mathrm{tot}}$
increases to more than~10. (Contrary to ordinary hydrodynamics,
$r_{\mathrm{tot}}$ always significantly depends on the shock Mach number, 
even when the latter is very high, see e.g.~\citealt{Blasi2002a}.) 
The corresponding effective adiabatic index~$\gamma_{\mathrm{eff}}$ 
is plotted in the middle of the figure. As expected, it decreases from~5/3 
(thermal fluid) to~4/3 (relativistic fluid) or even below (as particles escape).
In initially poorly efficient accelerators (low~$\eta$), shock modifications are small, 
but steadily increase over time. On the other hand, in very efficient accelerators (high~$\eta$),
back reaction effects are very strong even at a very early stage, but then decrease over time.
In all cases, we see that parameters evolve rather slowly after the first hundred years.
These results are similar to those obtained by~\cite{Decourchelle2000a} in 1D,
assuming self-similarity of the hydrodynamic profiles and using a different acceleration scheme,
which cross-validates the models.

Finally, the evolution of the relative positions of the waves is shown at the bottom of the figure. 
Two main comments are in order. 
First, the region affected by the Rayleigh-Taylor instability 
does not seem to be affected by the acceleration of particles. 
Second, the forward shock can get very close to this turbulent region 
if the injection level is high enough (above $5\times10^{-4}$) 
--~the reverse shock also reaches the contact discontinuity,
but this is caused by the development of the Rayleigh-Taylor instability, 
independent of acceleration efficiency.
These observations agree with the results of~\cite{Blondin2001a} and of~\cite{Fraschetti2009a}, 
obtained with no space- and time-dependent model of acceleration.
Our simulations also show that the average distance between the forward shock 
and the contact discontinuity does not depend much on time in the case of efficient acceleration.
(Rayleigh-Taylor fingers grow steadily in time, but the forward shock moves away
as back reaction effects are progressively reduced.)

\subsection{Remnant morphology\label{sub:Morphology}}

The morphology of the remnant after 500~years is shown in Fig.~\ref{fig:map_onoff},
where we compare cases with (top half) and without (bottom half) back reaction of particles. 
We used here the injection recipe of \cite{Blasi2005a},
which gives the injection level $\eta$ as a function of $\xi$ 
(we adopt the common value $\xi=3.5$)
and of the subshock compression ratio $r_{\mathrm{sub}}$ 
(so that injection is switched off as the shock gets too modified). 
Although $\eta$ is time-dependent, it remains close to the same value
during the whole simulation, going down from $4.6\times 10^{-4}$ to $3.9\times 10^{-4}$.

On the left of the figure, we show slices of the density, which highlights the remnant's structure.
The main effect of particle back reaction is apparent, namely that the shocked region shrinks. 
For this level of injection, this causes density to rise by a factor typically of two 
downstream of the forward shock.
Nevertheless, the size of the region perturbed by the Rayleigh-Taylor instability, 
the development of which depends on the density contrast, is basically unchanged.

On the right of the figure we show projected maps of the square of the density of shocked ejecta,
which roughly approximates the intensity of their thermal X-ray emission.
The interior of the remnant is filled with a filamentary texture, 
structured over scales compatible with the picture of \cite{Warren2005a}.
Even in projection, the unstable region is still clearly visible as a bright annulus.
But here again, cases with and without efficient acceleration 
cannot be distinguished from the structure of the shocked ejecta alone.

\subsection{Application\label{sub:Application}}

Although the detailed modelling of a specific object is beyond the scope of this Letter,
we can show the interest of our approach by comparing 
our numerical results with the observational results of \cite{Warren2005a}
regarding \emph{Tycho}'s remnant (see also \citealt{Volk2008b} for a study
of this object using a 1D non-linear kinetic model). 
Their main finding is that the forward shock is very close 
to the contact discontinuity, with a mean FS:CD ratio of 1:0.93. 
Assuming a uniform medium ($s=0$, a reasonable hypothesis for \emph{Tycho}), 
the theoretical ratio is 1:0.85 for an ejecta indice $n=7$ 
(assuming a self-similar evolution without particle acceleration).
As $n$ increases, the predicted ratio decreases, but is still only 1:0.89 for $n=14$.
As pointed out by \cite{Dwarkadas1998a}, an exponential profile 
may be more adapted to the early stages of an SN~Ia; however,
this would produce an even broader shocked region at the time considered \citep{Cassam-Chenai2008a}.
On the other hand, such a high FS:CD ratio can be obtained readily with our code
through the back reaction of accelerated particles. The procedure used by \cite{Warren2005a}
seems to extract the envelope of the ejecta, which can be obtained in our simulations
by setting some threshold on the ejecta fraction - we found the value of 10\% used here to give comparable results. 
Then the simulations can reproduce the observations provided that 
$\eta$ is of the order of $5\times 10^{-4}$ (\citealt{Volk2008b} obtained $\eta=3\times 10^{-4}$).
Although we have not conducted a complete parametric study yet, we believe 
that uncertainties in SNR parameters or in observed ratios could 
be accounted for by varying the injection level --~and reciprocally,
a good knowledge of the object will allow this crucial parameter to be constrained.

\cite{Warren2005a} also found that that the reverse shock is deep inside the ejecta, 
with a mean FS:RS ratio of 1:0.71, which is quite puzzling. 
For $n=7$ and $s=0$ the predicted ratio (without particle acceleration) is 1:0.79.
Projection effects might cause underestimates of the radius of the reverse shock, 
but according to our maps this effect is too small to explain the discrepancy. 
Considering more realistic profiles from explosion models might help understanding the situation.
In any case, this observation does not favour efficient acceleration at the reverse shock, 
which would shrink the whole remnant's structure even more.
However, other potentially important effects
(like the composition and temperature of the ejecta)
have to be included in our model before drawing firm conclusions.

\section{Conclusion}
\label{sec:Conclusion}

We have presented a new code that couples 
a 3D hydrodynamic description of a supernova remnant, 
allowing consideration of hydrodynamic instabilities such as the Rayleigh-Taylor instability, 
with a realistic kinetic model of non-linear acceleration at the blast wave,
allowing evaluation of the efficiency of particle acceleration.
We are thus now able for the first time to simulate the morphology of SNRs 
undergoing efficient DSA without limiting assumptions on its spatial structure or temporal evolution.

Our first results confirm the most notable previous findings 
regarding particle back reaction on the SNR morphology:
(i)~the shock structure is all the more compact since acceleration is efficient,
which provides a clear observational diagnostic; and (ii)~the development 
of the Rayleigh-Taylor instability is not significantly affected by acceleration at the forward shock,
but it has to be taken into account when interpreting observations.

Regarding the case of \emph{Tycho}'s remnant, comparison of our simulations
with X-ray observations strengthens the case for efficient acceleration of protons
at the forward shock.


\section*{Acknowledgements}
This work has been partially funded by the ACCELRSN project ANR-07-JCJC-0008.
The numerical simulations were performed with the DAPHPC cluster at CEA/Irfu.
The authors thank the anonymous referee for his/her useful comments.

\bibliographystyle{aa}
\bibliography{13666}

\begin{thebibliography}{27}
\expandafter\ifx\csname natexlab\endcsname\relax\def\natexlab#1{#1}\fi

\bibitem[{Amato \& Blasi(2006)}]{Amato2006a}
Amato, E. \& Blasi, P. 2006, \mnras, 371, 1251

\bibitem[{Berezhko \& Ellison(1999)}]{Berezhko1999a}
Berezhko, E.~G. \& Ellison, D.~C. 1999, \apj, 526, 385

\bibitem[{Blasi(2002)}]{Blasi2002a}
Blasi, P. 2002, Astroparticle Physics, 16, 429

\bibitem[{Blasi(2004)}]{Blasi2004a}
Blasi, P. 2004, Astroparticle Physics, 21, 45

\bibitem[{Blasi {et~al.}(2005)Blasi, Gabici, \& Vannoni}]{Blasi2005a}
Blasi, P., Gabici, S., \& Vannoni, G. 2005, \mnras, 361, 907

\bibitem[{Blondin \& Ellison(2001)}]{Blondin2001a}
Blondin, J.~M. \& Ellison, D.~C. 2001, \apj, 560, 244

\bibitem[{Cassam-Chena{\"\i} {et~al.}(2008)Cassam-Chena{\"\i}, Hughes, Reynoso,
  Badenes, \& Moffett}]{Cassam-Chenai2008a}
Cassam-Chena{\"\i}, G., Hughes, J.~P., Reynoso, E.~M., Badenes, C., \& Moffett,
  D. 2008, \apj, 680, 1180

\bibitem[{Chevalier(1983)}]{Chevalier1983a}
Chevalier, R.~A. 1983, \apj, 272, 765

\bibitem[{Chevalier {et~al.}(1992)Chevalier, Blondin, \&
  Emmering}]{Chevalier1992a}
Chevalier, R.~A., Blondin, J.~M., \& Emmering, R.~T. 1992, \apj, 392, 118

\bibitem[{Decourchelle {et~al.}(2000)Decourchelle, Ellison, \&
  Ballet}]{Decourchelle2000a}
Decourchelle, A., Ellison, D.~C., \& Ballet, J. 2000, \apjl, 543, L57

\bibitem[{Dwarkadas(2000)}]{Dwarkadas2000a}
Dwarkadas, V.~V. 2000, \apj, 541, 418

\bibitem[{Dwarkadas \& Chevalier(1998)}]{Dwarkadas1998a}
Dwarkadas, V.~V. \& Chevalier, R.~A. 1998, \apj, 497, 807

\bibitem[{Ellison {et~al.}(2004)Ellison, Decourchelle, \&
  Ballet}]{Ellison2004a}
Ellison, D.~C., Decourchelle, A., \& Ballet, J. 2004, \aap, 413, 189

\bibitem[{Ellison {et~al.}(2005)Ellison, Decourchelle, \&
  Ballet}]{Ellison2005a}
Ellison, D.~C., Decourchelle, A., \& Ballet, J. 2005, \aap, 429, 569

\bibitem[{Ellison {et~al.}(2007)Ellison, Patnaude, Slane, Blasi, \&
  Gabici}]{Ellison2007a}
Ellison, D.~C., Patnaude, D.~J., Slane, P., Blasi, P., \& Gabici, S. 2007,
  \apj, 661, 879

\bibitem[{Fraschetti {et~al.}(2009)Fraschetti, Teyssier, Ballet, \&
  Decourchelle}]{Fraschetti2009a}
Fraschetti, F., Teyssier, R., Ballet, J., \& Decourchelle, A. 2009, \aap
  (submitted)

\bibitem[{Gabici(2008)}]{Gabici2008a}
Gabici, S. 2008, in Proceedings of the XXI European Cosmic Ray Symposium

\bibitem[{Jones \& Ellison(1991)}]{Jones1991a}
Jones, F.~C. \& Ellison, D.~C. 1991, Space Science Reviews, 58, 259

\bibitem[{Malkov \& Drury(2001)}]{Malkov2001c}
Malkov, M.~A. \& Drury, L.~O. 2001, Reports on Progress in Physics, 64, 429

\bibitem[{Miceli {et~al.}(2009)Miceli, Bocchino, Lakubovskyi, Orlando,
  Telezhinsky, Kirsch, Petruk, Dubner, \& Castelletti}]{Miceli2009a}
Miceli, M., Bocchino, F., Lakubovskyi, D., {et~al.} 2009, \aap, 501, 239

\bibitem[{Patnaude \& Fesen(2009)}]{Patnaude2009b}
Patnaude, D.~J. \& Fesen, R.~A. 2009, \apj, 697, 535

\bibitem[{Teyssier(2002)}]{Teyssier2002a}
Teyssier, R. 2002, \aap, 385, 337

\bibitem[{Truelove \& McKee(1999)}]{Truelove1999a}
Truelove, J.~K. \& McKee, C.~F. 1999, \apjs, 120, 299

\bibitem[{Velazquez {et~al.}(1998)Velazquez, Gomez, Dubner, de~Castro, \&
  Costa}]{Velazquez1998a}
Velazquez, P.~F., Gomez, D.~O., Dubner, G.~M., de~Castro, G.~G., \& Costa, A.
  1998, \aap, 334, 1060

\bibitem[{V{\"o}lk {et~al.}(2008)V{\"o}lk, Berezhko, \&
  Ksenofontov}]{Volk2008b}
V{\"o}lk, H.~J., Berezhko, E.~G., \& Ksenofontov, L.~T. 2008, \aap, 483, 529

\bibitem[{Wang \& Chevalier(2001)}]{Wang2001a}
Wang, C.-Y. \& Chevalier, R.~A. 2001, \apj, 549, 1119

\bibitem[{Warren {et~al.}(2005)Warren, Hughes, Badenes, Ghavamian, McKee,
  Moffett, Plucinsky, Rakowski, Reynoso, \& Slane}]{Warren2005a}
Warren, J.~S., Hughes, J.~P., Badenes, C., {et~al.} 2005, \apj, 634, 376

\end{thebibliography}

\end{document}